# Optical continuum structure of Cygnus A

R.A.E. Fosbury[1], J. Vernet[1], M. Villar-Martín[2],
M.H.Cohen[3], P.M. Ogle[3], H.D. Tran[4] & R.N. Hook[1]

1. ST-ECF, ESO, Garching bei München, Germany (rfosbury@eso.org)
2. Department of Physics, University of Sheffield, UK
3. Astronomy Department, California Institute of Technology, Pasadena, USA
4. IGGP/LLNL, Livermore, USA

**Abstract**

As a prerequisite for interpreting new observations of the most distant radio galaxies, we make an optical study of the closest powerful radio source, Cygnus A. Using Keck imaging- and spectro-polarimetry in conjunction with HST, WFPC2 broad and narrow band imaging, we are able to identify specific geometrical structures in the galaxy with optical continuum components distinguished by both colour and polarimetric properties. A 4 kpc diameter dusty ring of young stars forms the equator of a double ionization cone which is co-axial with the radio jets.

## 1 Introduction

Isolating and identifying the emitting components of the most powerful radio galaxy in the local universe is a prerequisite for analysing the properties of the most distant members of the class. Cygnus A, with a redshift of only 0.056 and first studied some 43 years ago by Baade & Minkowski (1954), has an optical structure which has proved extraordinarily difficult to disentangle. There is now a large literature, best accessed by reading the review article by Carilli & Barthel (1996) and the proceedings of the recent NRAO workshop (Carilli & Harris, 1996). The long-known presence of a very strong emission line spectrum covering a wide range of ionization (Osterbrock & Miller 1975) and the detection of weak optical polarization (Tadhunter, Scarrott & Rolph 1990) gave some support to the notion of a buried quasar, but there was always concern that such a nucleus should betray itself in a more obvious fashion. The realisation that Cygnus A has an unusually low FIR to radio luminosity ratio and the suggestion that the object may be overluminous in the radio — due to its dense environment — rather than underluminous at other wavelengths (Barthel & Arnaud 1996) has made people more willing to accept the (weak) quasar hypothesis, but perhaps less willing to regard the object as a good prototype for the powerful sources at high redshift.

Weak though it may be, the presence of a radio quasar in a galaxy at such a low redshift gives us a unique opportunity to take it apart and catalogue the properties of those components which, albeit in different proportions, may be used to build the high redshift objects.

One of the reasons for the relatively slow progress in understanding this object is that it lies behind a veil of Galactic dust ($A_V \approx 1.5$) and its nuclear regions are further obscured by local dust and seen against the emission from an elliptical galaxy which contributes more than half of the visible light. This has meant that more powerful observational tools have been required than might have been expected for an object of this intrinsic luminosity and distance.

In spite of extensive earlier efforts, it took HST FOS spectroscopy (Antonucci, Hurt & Kinney 1994) and the Keck spectropolarimeter (Ogle et al. 1997, hereafter O97) to find the scattered broad line spectrum of Mg II and H$\alpha$. The Balmer lines are unusually broad (FWHM $\approx$ 26,000 km s$^{-1}$) for quasars, resembling some of the broad line radio galaxies. This extent of BLR light in wavelength undoubtedly contributed to previous non-detections. In addition to polarized, broad H$\alpha$, O97 reported weakly polarized narrow lines redshifted by 110–230 km s$^{-1}$.

The new work reported here is an analysis of the spatial variations seen in the Keck imaging- and spectro-polarimetry, interpreted in combination with multicolour images taken with WFPC2 on HST (Jackson & Tadhunter, in prep.) and available to us in the public archive. The clear identification of the spectral with the spatial components which results from the application of these two observational techniques demonstrates the extaordinary power of this combination of tools.

## 2 Observations

The imaging- and spectropolarimetric observations were obtained with the Low-Resolution Imaging Spectrograph polarimeter (LRISp, Oke et al. 1995; Goodrich, Cohen & Putney 1995) on the Keck II telescope during October 1996. The data and their reduction are described in O97. The result of one hour of imaging polarimetry in the $B$-band is shown in Figure 1 that has the spectrograph slit and the extraction apertures overlayed. 2.2 hours of spectropolarimetric data were obtained in the wavelength range 3900–8900Å with a slit of width 1 arcsec in PA 101°. The spectra have been corrected for a Galactic reddening of $E_{B-V} = 0.5$ and an elliptical galaxy template (NGC 821) has been subtracted from the spatially extracted spectra. The elliptical galaxy at 5500Å contributes 64% of the flux in the east and 62% in the western extractions.

The HST WFPC2 images, from programmes by Westphal and by Jackson, were retrieved from the public archive at the ST-ECF. Exposures with the filters

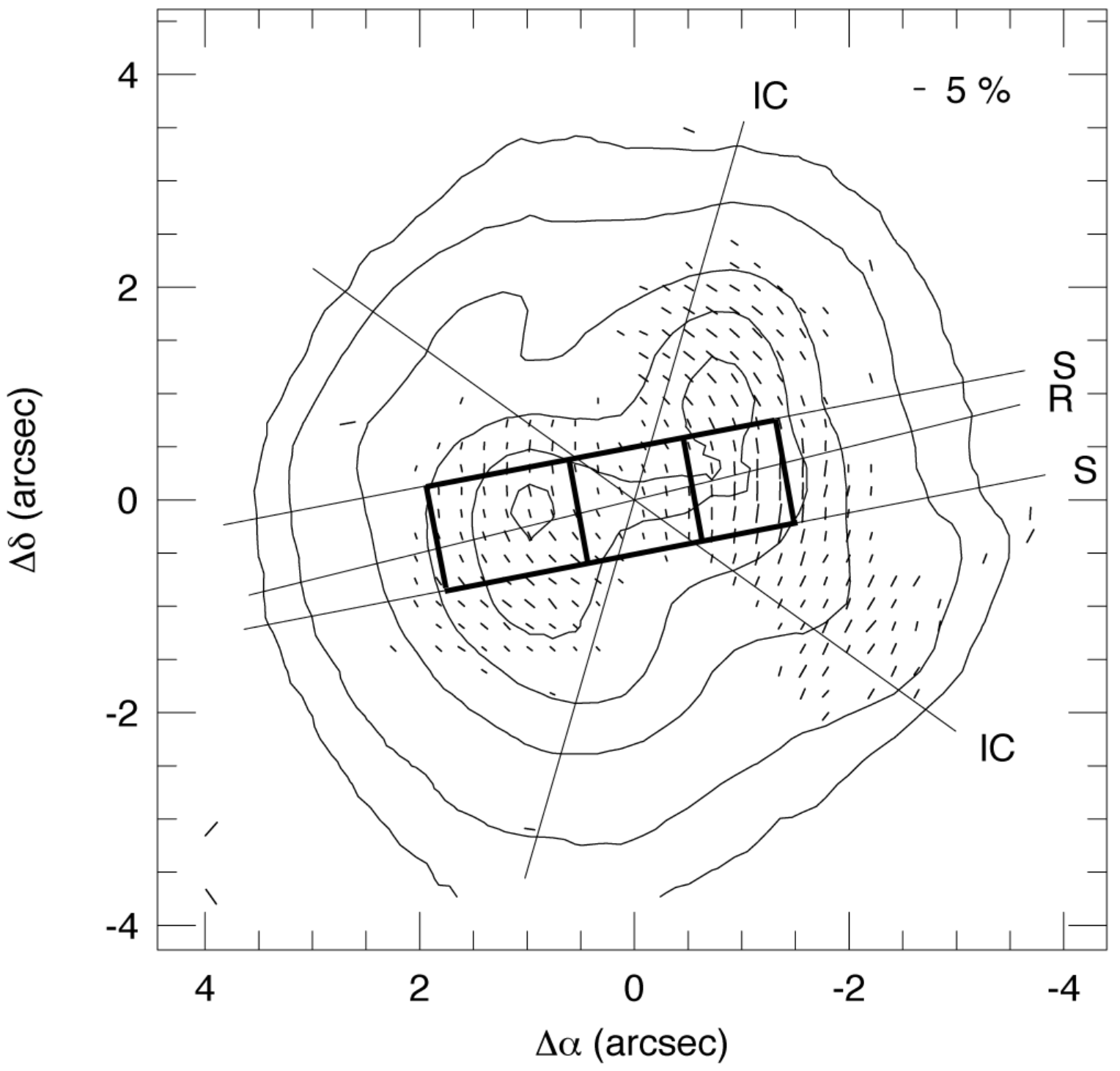

**Figure 1.** Keck II $B$-band imaging polarimetry of Cygnus A. $E$-vectors with SNR≥2.5 are shown as short lines on the intensity contour map. The spectrograph slit (S) together with the boxes showing the three extraction apertures, the radio axis (R) and the ionization cone (IC) from the HST images (Jackson et al. 1996) are marked.

F450W, F550W, F622W, F814W and narrow band sub-frames using the linear ramp filters at the wavelengths of the redshifted [O III], H$\alpha$ and [O I] lines were cleaned of cosmic ray events and co-aligned using stellar images. The F450W image, which has low signal/noise ratio due to the large Galactic extinction in Cygnus, has been co-added with the combined $B$-band images from LRISp using the ACOADD task in IRAF developed by Hook & Lucy (1993).

## 3 Discussion

In this discussion, we focus principally on the nature of the continuum that remains after the subtraction of the red population of stars in the elliptical galaxy. From the spectropolarimetry integrated along the central 7.6 arcsec of the slit, O97 were able to use the variation of $P$ and P.A. with wavelength to model the continuum with three components: a nebular continuum computed from the observed, Galactic extinction corrected narrow H$\beta$ flux, a blue ($f_\nu \propto \nu^{+2}$) featureless continuum (FC1) polarized similarly to the broad H$\alpha$, and a second continuum (FC) which is redder ($f_\nu \propto \nu^{-1}$) and has lower polarization in a different P.A. FC1 was attributed to dust scattering of a hidden AGN of rather moderate luminosity. The nature of the second FC was undetermined although it was suggested that it could be slightly polarized radiation from dichroically absorbed or scattered light from OB stars associated with the blue, knotty structures seen in the HST images by Jackson et al. (1996).

The very different behaviour of the total and polarized flux continua in the eastern and western spectral extractions was pointed out by O97. The polarized flux is bluer than the total flux in the western component, but the opposite is observed in the east. Here we attempt to fit the continua in these two regions using the minimum number of components required to explain the polarization behaviour. By using colour images constructed from the HST and Keck filter data, we are able to make a direct association between geometrically identifiable structures in the galaxy and the continuum components necessary to synthesize the spectropolarimetry.

The most revealing colour maps are those constructed from the broad band $B$(F450W), F622W and F814W filters, which represent predominantly continuum radiation (F555W is not used because it contains [O III]), and the $B$(F450W), [O III] and H$\alpha$ filters. These are shown as the two panels in Plate 4 where the three filter images are assigned to the blue, green and red channels respectively.

The line radiation delineates a symmetric biconical structure with an axis close to that of the radio jets whereas the blue continuum extends well outside the cone in the east and appears to be in an equatorial ring which partially obscures the eastern cone and is obscured by the western one. The continuum image reveals extensive dust structures that appear to be loosely associated with the blue ring in a way reminiscent of Centaurus A. The continuum image also shows a very red (presumably) Galactic star that falls within the western aperture and is accounted for in the spectral fitting process. It is clear that the strong blue continuum in the east noted by several previous authors (see Carilli & Barthel 1996) arises from the apparent ring structure that appears not to be associated directly with the ionization cones.

| East | | | | |
|---|---|---|---|---|
| Comp. | P(%) | PA(°) | $F/F_{6563}$ | $E_{B-V}$ |
| $FC1_A$ | 43 | 48 | .20 | 1.2 |
| $FC1_B$ | 25 | 18 | .15 | 0 |
| FC2 | 0 | - | .65 | 0 |
| **West** | | | | |
| Comp. | P(%) | PA(°) | $F/F_{6563}$ | $E_{B-V}$ |
| $FC1_A$ | 23 | -1 | .40 | 0 |
| $FC1_B$ | 15 | -3 | .40 | 1.2 |
| Red Star | 0 | - | .20 | 0 |

**Table 1.** Parameters for the fitted continuum components in the eastern and western extractions.

We have fitted the spectropolarimetric continuum data separately in the eastern and western extractions with the components listed in Table 1. The fits themselves are shown in Figure 2.

In both cases, the nebular continuum has been calculated from the observed H$\beta$ flux and has been given the (low) $P$ and P.A. measured for the narrow emission lines. In the west, the continuum over most of the observed spectral range is dominated by a polarized FC1, represented everywhere by power law with $f_\nu \propto \nu^{+2}$, with low reddening. The polarization is diluted in the red by both the nebular continuum and the Galactic star ($T_{eff} = 3,500$K). The fit to the polarimetry is improved slightly by the addition of a second FC1 component with a much larger reddening and a slightly different P.A. As there are several spatially separated knots of continuum radiation visible in the HST images within the western aperture, the presence of more than one FC1 component is not surprising.

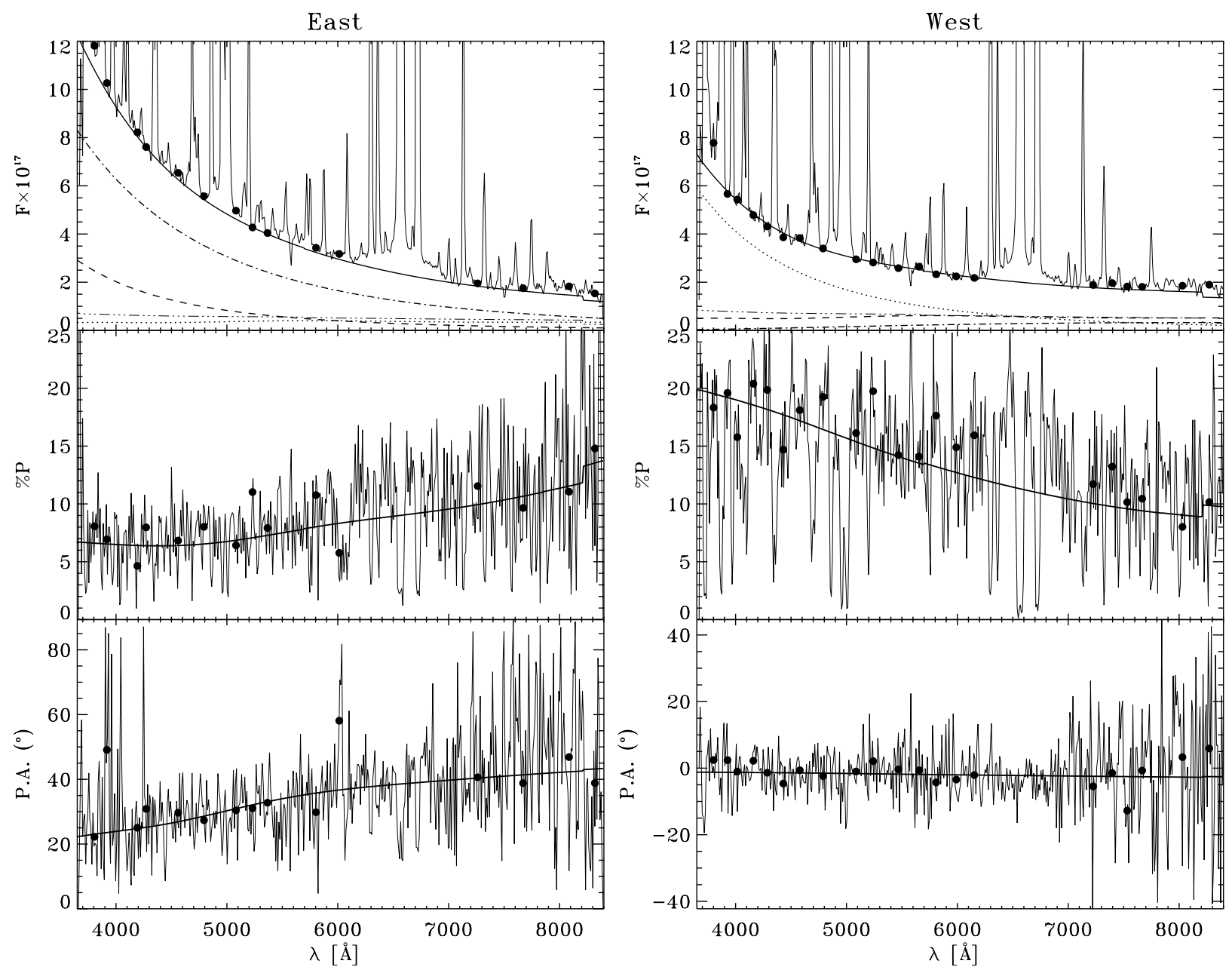

**Figure 2.** Continuum fits for the eastern and western spectropolarimetric data (see Table 1). The top panel shows the total flux spectrum with binned continuum points as filled circles. The continuum components are: ...... $FC1_A$; - - - - $FC1_B$; -.-.-. FC2/red star; -...-...- Nebular continuum. Although the $P$ and P.A. spectra are plotted in the lower two panels, the fits were made to the continuum bins in the $I$, $q$ and $u$ spectra.

The spectrum in the east is dominated by a very blue unpolarized continuum which we call FC2 (Tran 1995, Tran, Cohen & Goodrich 1995). This is represented by a black body with $T_{eff} = 25,000$K. The polarization behaviour is then fitted by adding two FC1 components, one with significant reddening and a P.A. of 48° and another with no reddening and a P.A. of 18°. This is justified, as Figure 1 shows, because the $E$-vectors within the eastern aperture, unlike those in the west which have an almost constant orientation, range from around 10° to 50°. Evidence that the unpolarized FC2 consists of blue starlight is provided by the behaviour of the Balmer emission lines within the eastern and western apertures. Figure 3 shows the ratio of prominent emission lines spanning the whole wavelength range of the data from the east and west. With the exception of the blue [Ne III] line and the Balmer lines, most emission lines indicate that the NLR is more reddened in the east by $\Delta E_{B-V} = 0.07$. The anomalous behaviour of the Balmer lines can be explained if FC2 is not entirely featureless but exhibits Balmer absorption with equivalent widths ranging from about 5Å for H$\delta$ to 14Å for H$\beta$ which will reduce the measured emission line fluxes in the east.

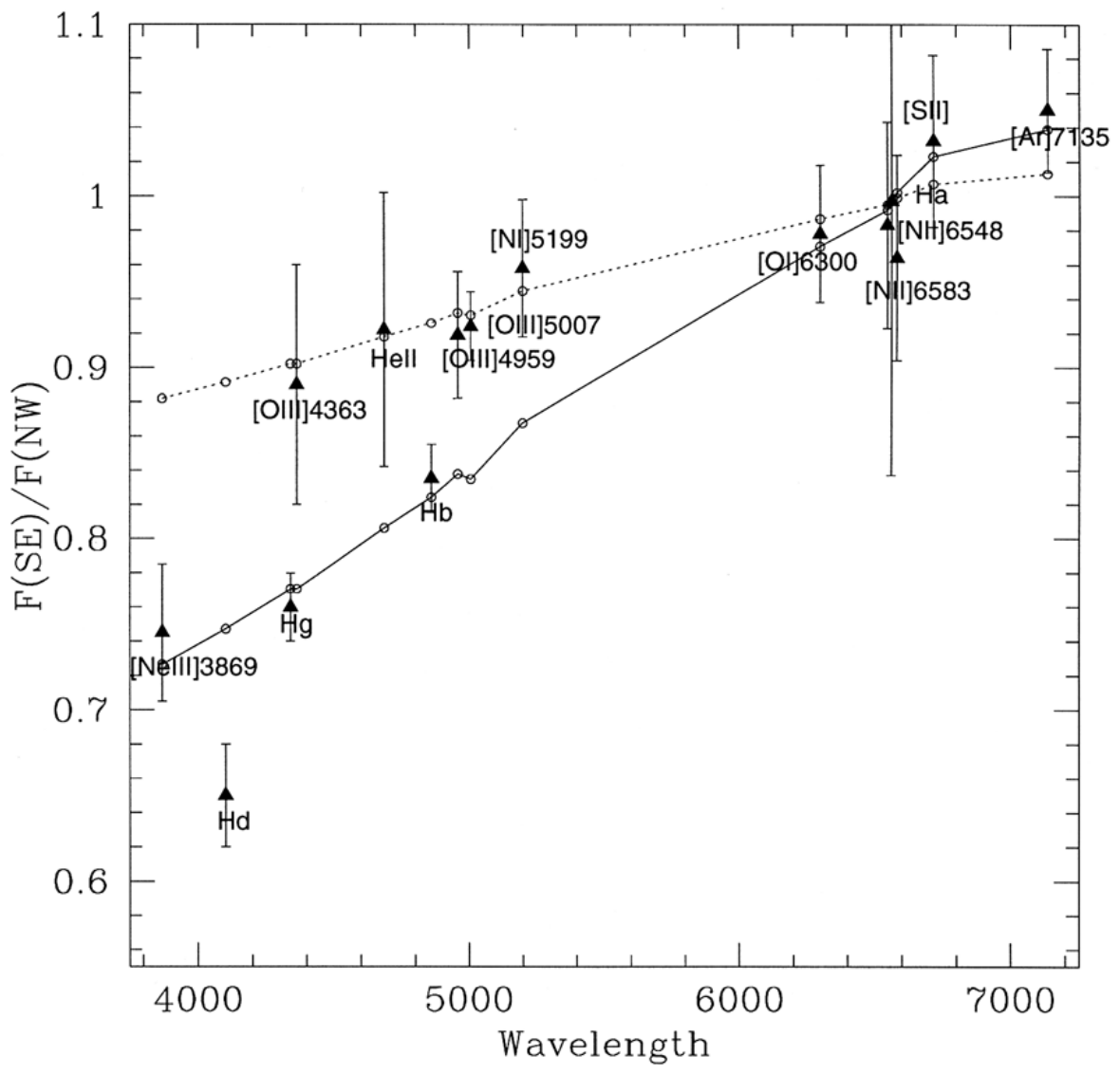

**Figure 3.** The ratio of emission line intensities from the eastern and western apertures plotted as a function of wavelength. Most of the lines are consistent with an excess reddening in the east with $\Delta E_{B-V} = 0.07$ shown as the dashed line: the solid line represents $\Delta E_{B-V} = 0.17$. The anomalous behaviour of the Balmer lines is due to underlying hydrogen absorption lines in the spectra of the hot stars which comprise the FC2.

# 4 Conclusions

The picture that we now have — from O97, this work and the previous literature — of the optical emission from the Cygnus A galaxy is represented as the cartoon

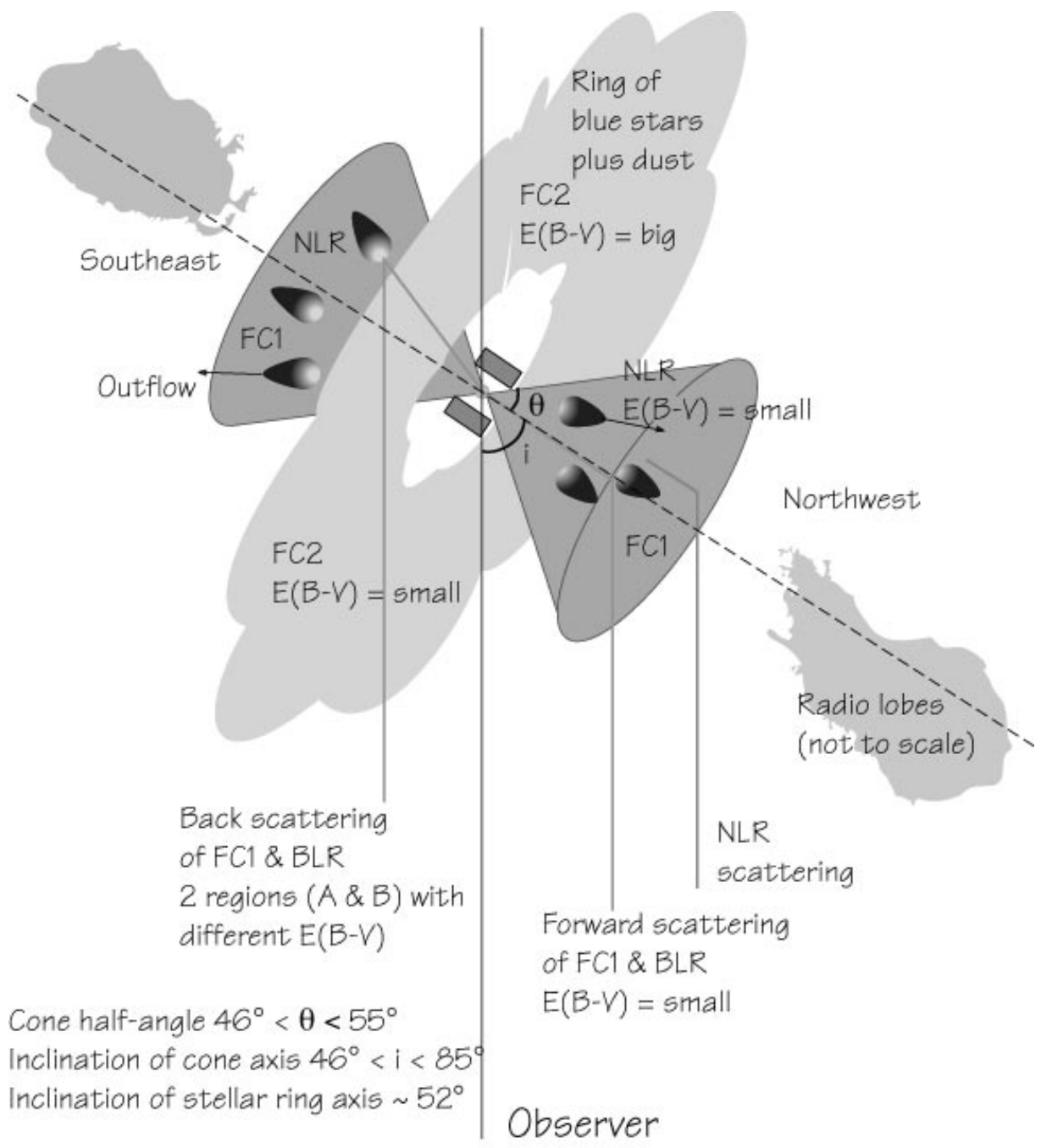

**Figure 4.** A cartoon representing the geometric configuration of the spectral components identified from the spectropolarimetry and the colour images. The cone angle and the inclination axis are discussed in O97. The orientation of the stellar ring axis is derived from its apparent axial ratio.

in Figure 4.

A dust-scattered AGN continuum from a quasar or broad line radio galaxy of modest luminosity is confined to an ionization bicone. This FC1 is highly polarized (locally up to nearly 50%) with the $E$-vector perpendicular to the line connecting it to the nucleus. Its spectral index is blue, $\alpha = +2$ ($F_\nu \propto \nu^\alpha$), suggesting that the scattering process bluens the AGN continuum by $\Delta\alpha \approx 3-4$. Reddening differences suggest that the western cone is on the nearside of the galaxy, consistent with the radio jet data (Carilli & Barthel 1996) if the cone and the jets are co-axial.

The blue unpolarized continuum (FC2) seen in the east is from a population of young stars, represented by $T_{eff} = 25,000$K in our fits. This, together with the HST colour images, suggests the stars reside in a dusty equatorial ring with a diameter of around 4 kpc and an axis coinciding with the radio and ionization cone axes. The young stars avoid the region of the radio jets; they are in an orthogonal structure.

There is a nebular continuum associated with the regions emitting the NLR. This contributes significantly only in the red. There are polarized, dust scattered, broad permitted lines: H$\alpha$ and H$\beta$ in our data and Mg II in the HST spectroscopy of Antonucci, Hurt & Kinney (1994).

Narrow emission lines are emitted mostly within the ionization cones. These are more reddened in the east and are weakly polarized in a manner suggestive of scattering in an outflowing wind of a few hundred km $s^{-1}$.

All of these components are seen in high redshift radio galaxies. Our view of the aligned light in these objects is clearer because the redshift moves our observation window below the spectral region dominated by the red stars in the elliptical galaxy. The balance between the FC1 and FC2 luminosity clearly varies from object to object and might be expected to evolve in a particular object. We do not yet, however, have a very clear picture of the relative spatial distribution of these two blue continua in the more distant objects. It is noteworthy, however, that in this one nearby example of a powerful radio source, the scattered FC1 comes from along the radio axis while the stellar FC2 does not. There appears to be no recent star formation associated with the jets.

**Acknowledgements** We thank Neal Jackson for discussions regarding the HST data. The HST images were obtained from the public HST archive operated by the ST-ECF. The W.M. Keck Observatory is operated as a scientific partnership between the California Institute of Technology, the University of California and the National Aeronautics and Space Administration. RAEF is affiliated to the Astronomy Division, Space Science Department, European Space Agency.

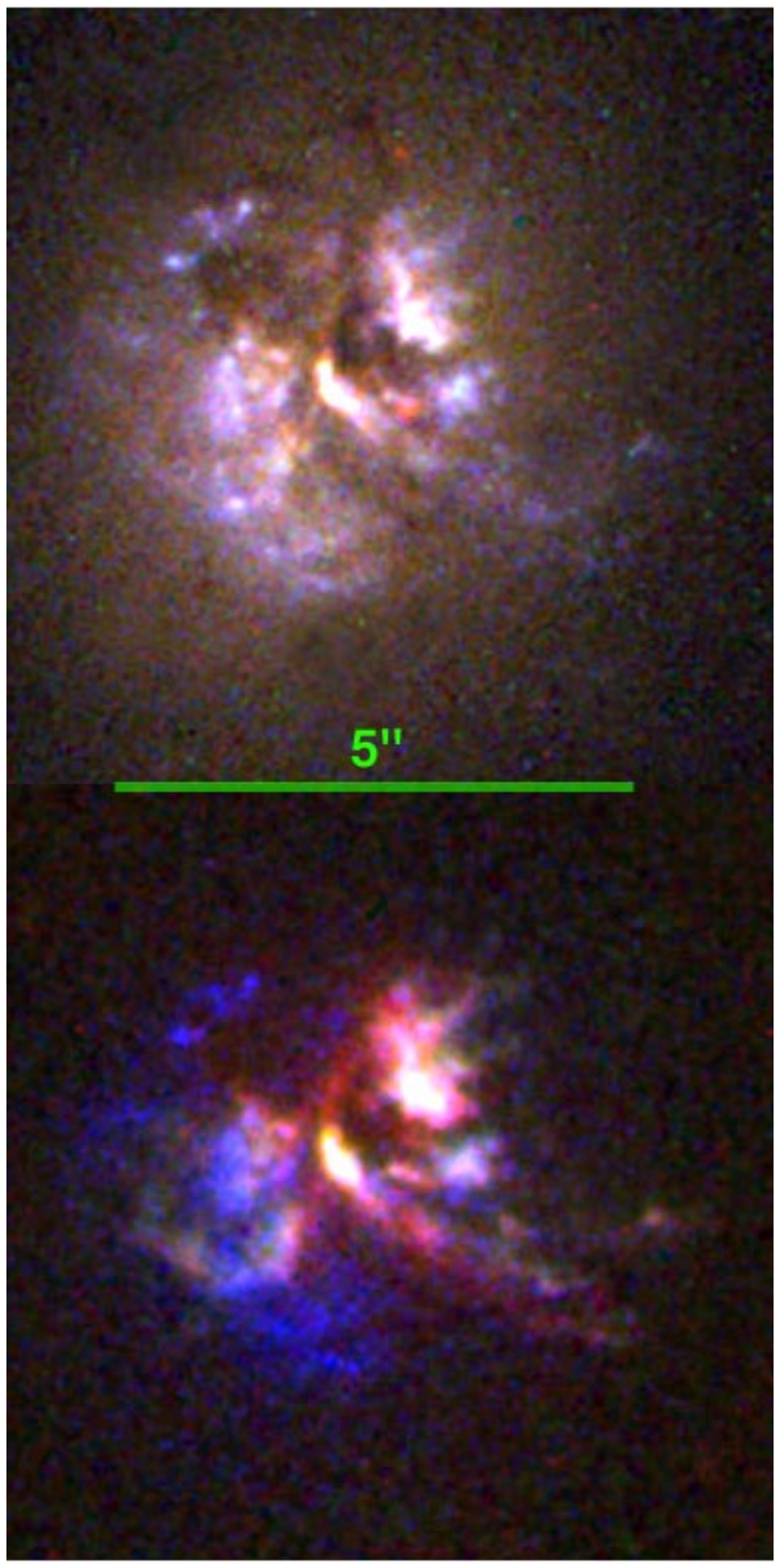

**Plate 4.** Colour images of Cygnus A. (top) 'True' colour continuum image constructed from the Keck $B$-band and the HST F450W combined image (blue), the HST F622W (green) and HST F814W (red) data. There is some line contamination in these filters but the very strong [O III] and H$\alpha$ radiation is largely excluded. (bottom) The $B$(F450W) (blue), [O III] (green) and H$\alpha$ (red) composite shows the very marked difference between the spatial distribution of the blue continuum and the strong emission lines.